\documentstyle[12pt]{article}
\newcommand{\be}{\begin{equation}}
\newcommand{\ee}{\end{equation}}

\begin{document}
\renewcommand{\baselinestretch}{1.4}
\small\normalsize

\begin{flushright}
{\sc BUHEP}-96-42\\September 1996
\end{flushright}

\vspace*{1.5cm}

\begin{center}

{\Large \bf Critical properties and monopoles\\in U(1) lattice gauge theory}

\vspace*{0.8cm}

{\bf Werner Kerler$^a$, Claudio Rebbi$^b$, Andreas Weber$^a$}

\vspace*{0.3cm}

{\sl $^a$ Fachbereich Physik, Universit\"at Marburg, D-35032 Marburg,
Germany\\
$^b$ Department of Physics, Boston University, Boston, MA 02215, USA}
\hspace*{3.6mm}

\end{center}

\vspace*{1.5cm}

\begin{abstract}
We present a detailed study of the properties of the phase transition
in the four-dimensional compact U(1) lattice gauge theory supplemented
by a monopole term, for values of the monopole coupling $\lambda$ such
that the transition is of second order. By a finite size analysis
we show that at $\lambda= 0.9$ the critical exponent is already
characteristic of a second-order transition. Moreover, we find
that this exponent is definitely different from the one of the
Gaussian case.  We further observe that the monopole density becomes
approximately constant in the second-order region. Finally we reveal
the unexpected phenomenon that the phase transition persists up
to very large values of $\lambda$, where the transition moves to
(large) negative $\beta$.
\end{abstract}

\newpage

\section{Introduction}

\hspace{3mm}

The continuum limit can be recovered from the lattice regularization
of a quantum field theory by approaching a critical point with
suitable scaling properties.  While the existence of such a critical
point can be established through perturbative considerations in the
case of asymptotically free theories, very little is known about
critical points that can give origin to continuum limits in the
non-perturbative domain.  The $U(1)$ lattice gauge theory, although
characterized by the most elementary continuous gauge symmetry, has a
rather complex phase structure, induced by the presence of monopole
excitations.  While QED, which shares the same gauge group, appears
very successfully embedded in the $SU(2)\times U(1)$ theory of
electroweak interactions, it is still quite interesting to study
the properties of the pure $U(1)$ lattice gauge theory in view of
what one can learn from the existence of a non-trivial critical point
and its implications for possible continuum limits of lattice gauge
theories in the non-perturbative domain.

We have devoted several investigations to the study of the properties
of the $U(1)$ lattice gauge theory, taking special advantage of a technique 
we introduced to bridge the valley between energy gaps \cite{krw94,krw95a}, 
and of a new order parameter based on the topological properties of monopole 
networks\cite{krw94,krw95}. Our investigations have been based
on the study of the model with the Wilson action supplemented by an
additional monopole term \cite{bs85}

\be
S=\beta \sum_{\mu>\nu,x} (1-\cos \Theta_{\mu\nu,x})+
\lambda \sum_{\rho,x} |M_{\rho,x}|
\label{sbl}
\ee
where $M_{\rho,x}=\epsilon_{\rho\sigma\mu\nu}
(\bar{\Theta}_{\mu\nu,x+\sigma}-\bar{\Theta}_{\mu\nu,x}) /4\pi$
and the physical flux $\bar{\Theta}_{\mu\nu,x}\in [-\pi,\pi)$ is related
to the plaquette angle $\Theta_{\mu\nu,x}\in (-4\pi,4\pi)$ by
$\Theta_{\mu\nu,x}=\bar{\Theta}_{\mu\nu,x}+2\pi n_{\mu\nu,x}$ \cite{dt80}.

It is well known that this system undergoes a phase transition at a
coupling $\beta_{\mbox{\scriptsize cr}}(\lambda)$ separating a
disordered phase, characterized by a condensate of monopole
excitations, for $\beta<\beta_{\mbox{\scriptsize cr}}$, from an
ordered, Coulomb-like phase for $\beta>\beta_{\mbox{\scriptsize cr}}$.
Our own studies of the energy distribution have shown that as
$\lambda$ increases the phase transition changes nature: a gap in the
energy distribution indicates that the transition is of the first
order for $\lambda=0$ and for moderately small values of $\lambda$,
whereas the disappearance of the gap for larger values of $\lambda$
indicates that the transition becomes there of the second order. In
this paper we wish to contribute further to the knowledge of the
$U(1)$ lattice gauge theory by exploring in detail the region of
larger $\lambda$ couplings, where the phase transition appears to be
continuous.  An important aim of our investigation is to verify by a
finite size analysis that the transition is indeed of second order and
to determine deviations of the critical exponent from the Gaussian
case. A further goal is to clarify the behavior of the transition when
$\lambda$ becomes very large.  Because the phase transition is so closely
related to the appearance of a monopole condensate, exploring in
detail the effects of the monopole term in the action appears to be of
particular interest.

In this study we use periodic boundary conditions, guaranteeing
translational invariance and homogeneity.

\section{The line of phase transitions}

\hspace{3mm}
In the paper where they introduced the action (\ref{sbl}) Barber and
Shrock already made the observation that there is a shift of the
transition point if $\lambda$ is varied \cite{bs85}.  It has also been
reported in the literature \cite{bss85,bmm93} that a complete
suppression of monopoles, corresponding to $\lambda=\infty$, produces
the disappearance of the confining phase.  In our previous
investigations we determined the location of the phase transition
$\beta_{\mbox{\scriptsize cr}}(\lambda)$ for a wide range of values of
$\lambda$ \cite{krw94,krw95a}. Now we wish to clarify what happens at
very large $\lambda$. For this purpose we determine the critical
points at values of $\lambda$ substantially larger than previously
considered and if necessary, allow $\beta$ to become negative.

In order to keep the computational cost for this study within bearable
limits we have used our topological characterization of the phases
\cite{krw94,krw95} to locate the critical points. Our classification
is based on the fact that there is an infinite network of monopole
current lines in the confining phase and no such network in the
Coulomb phase.  On finite lattices ``infinite'' is to be defined in
accordance with the boundary conditions \cite{krw96}. For the periodic
boundary conditions we are considering here, ``infinite'' is
equivalent to ``topologically nontrivial in all directions''.  While
for loops the topological characterization is straightforward, to
determine the existence of an infinite network it is necessary to
perform a more elaborate analysis based on homotopy preserving
mappings \cite{krw94,krw95}.

Because the probability $P_{\mbox{\scriptsize net}}$ to find a network
which is nontrivial in all directions takes values exactly 1 and 0 in the
confining phase and in the Coulomb phase, respectively, the frequency
of the occurrence of such a network is a very advantageous order
parameter.  Indeed, with an infinite system a single configuration would
be sufficient to identify the phase.  On a finite lattice, because of
finite size effects, the distinction is not as sharp, but still few
configurations are sufficient to discriminate bewteen the phases.
Examples are given in Figure 1.

It should be remembered that on finite lattices different order
parameters lead to slightly different critical $\beta$. On an $8^4$
lattice the maximum of the specific heat and our topological order
parameter give values 1.0075(1) and 1.0074(2) for $\lambda=0$, and
0.3870(5) and 0.372(3) for $\lambda=0.9$, respectively. To determine
the location of the maximum of the specific heat for larger $\lambda$
in an efficient way we first determine the critical $\beta$ from the
topological order parameter and then find the maximum of the specific
heat in an easy second step.

In Figure 2 we show our results for the location of the phase
transition $\beta_{\mbox{\scriptsize cr}}$, defined by the maximum of
the specific heat, for values of $\lambda$ ranging up to $1.3$. It can
be seen that the phase transition line continues to negative $\beta$.

Using the topological order parameter we could follow the line of
phase transitions up to still much larger $\lambda$: from $\lambda =
1.4$ where $\beta_{\mbox{\scriptsize cr}} = -0.52(2)$ to $\lambda$ =
10 where $\beta_{\mbox{\scriptsize cr}}$ is approximately $-1000$. It
should be emphasized that the characteristic topological signatures
for the two phases are apparent throughout entire range of $\lambda$.
This shows that, quite remarkably, both phases persist all the way up
to very large $\lambda$.

We find that finite size effects increase with $\lambda$. This is
indicated by the fact that the transition region becomes broader. From
Figure 3 it can be seen that the width of the peak of the specific
heat increases and that its height decreases with increasing
$\lambda$. Obviously for very large $\lambda$ a precise determination
of the location of the maximum of the specific heat becomes quite
difficult. The increase of the width of the transition region with $\lambda$
is also apparent from the data on the topological order parameter
$P_{\mbox{\scriptsize net}}$ in Figure 1 (notice the different scales
for $\beta$ in the upper and lower parts of the figure).  Still, as we
already remarked above, the fact that the topological order parameter
takes values 0 and 1 in the two phases makes it easy to discriminate
between the two phases even at very large $\lambda$.

\section{Critical behavior}

\hspace{3mm}
In order corroborate our observation that for large $\lambda$ the
phase transition becomes of second order we have investigated the
finite-size scaling behavior of the maximum of the specific heat
$C_{\mbox{\scriptsize max}}$. It is expected to be
\be
C_{\mbox{\scriptsize max}} \sim L^d
\ee
if the phase transition is of first order and
\be
C_{\mbox{\scriptsize max}} \sim L^{\frac{\alpha}{\nu}}
\ee
if it is of second order, where $\alpha$ is the critical exponent of the
specific heat and $\nu$ the critical exponent of the correlation length.

In Figure 4 we present results of our simulations for
$C_{\mbox{\scriptsize max}}$, on lattices with $L$ = 6, 8, 10, 12 for
$\lambda = 0.9$ at the corresponding values of
$\beta_{\mbox{\scriptsize cr}}$.  The fit to these data gives
\be
{\frac{\alpha}{\nu}} = 0.485(35)
\ee
Clearly this is quite far from 4 and thus the transition not of first order.

 From the hyperscaling relation $\alpha = 2 - d\,\nu$ we find
\be
\nu = 0.446(5)
\label{nu}
\ee
It is interesting to note that (\ref{nu}) is clearly different from
the value $\frac{1}{2}$ of the Gaussian case.

The critical $\beta$ is expected to behave as
\be
\beta_{\mbox{\scriptsize cr}}(L)=\beta_{\mbox{\scriptsize
cr}}(\infty) +a L^{-\frac{1}{\nu}}
\ee
 From this relation, using the value in Eq.~(\ref{nu}) and our data for
$\beta_{\mbox{\scriptsize cr}}(L)$ at $\lambda = 0.9$, we get
$\beta_{\mbox{\scriptsize cr}}(\infty)$ = 0.4059(5) and $a = -1.99(6)$.

We have also performed simulations for $\lambda = 0.8$ on lattices with
$L$ = 6, 8, 10. It turns out that $C_{\mbox{\scriptsize max}}$
does not yet scale in this case. This might be related to the particular
sensitivity close to a tricritical point.

In Figure 5 we present the monopole density
$\rho= (1/4L^4)\sum_{\rho,x} |M_{\rho,x}|$ along the transition line.
It is interesting to observe that starting at $\lambda=0.9$ the density
becomes constant within the errors of the simulation.

\section{Discussion}

\hspace{3mm}
In the past values between 0.33 and 0.50 have been found for the critical
exponent $\nu$, in different contexts \cite{gnc86,l87}. This wide range of
values has prevented firm conclusions. More precise results, based on a
finite size analysis and higher statistics, have been recently reported
\cite{jln96} for the Wilson action extended by a double charge term with
coupling $\gamma$.

The well known features of this action are that the first order
transition, which occurs at $\gamma=0$, weakens with increasing
$\gamma$ until it becomes of second order at a tricritical
point. While for the usual periodic boundary conditions the second
order region starts at negative $\gamma$, the authors of
Ref.~\cite{jln96}, using a spherelike lattice, find evidence for a
second order transition already at $\gamma=0$.  Of course, this could
also be due to the fact that inhomogeneities weaken the transition
\cite{krw96} and that the mapping from the surface of a hypercube to
the 4-sphere can provide inhomogeneities.

The result $\nu=0.446(5)$ presented here and the result $\nu=0.365(8)$
given in Ref.~\cite{jln96} are both definitely different from the
value $\frac{1}{2}$ of the Gaussian case. Thus a continuum limit of
four-dimensional pure U(1) gauge theory, nontrivial and not
asymptotically free, appears possible.

The difference between the two results, though not dramatic, is too
large to be accounted for by systematic errors. Further studies appear
necessary. In particular, simulations at some value of $\lambda$
larger than 0.9 will be important for seeing whether the value of
Eq.~(\ref{nu}) is universal. Some of the results obtained in
Ref.~\cite{jln96} for three different values of $\gamma$ and using
different techniques, though close to each other, also differ beyond
the statistical errors.  Results for yet more negative $\gamma$ and
also for periodic boundary conditions would desirable.

The increase of the finite-size effects with lambda mentioned in
Sect.~2 could hint at an increase of $\nu$, related to a larger
correlation length $\eta \sim |\beta - \beta_{\mbox{\scriptsize
cr}}|^{-\nu}$.  The corresponding decrease of $\alpha$ from 0.216(16)
at $\lambda=0.9$ to $0$, in view of the relation $C \sim |\beta -
\beta_{\mbox{\scriptsize cr}}|^{-\alpha}$, also appears consistent
with the behavior of the specific heat seen in Figure 3. The behavior
of $\alpha$, in particular, lends support to the fact that the
increase of finite-size effects is due to the exponents and not to the
constants in front of the scaling expressions. Possibly $\nu$ could
reach in this way the Gaussian value.

In regard to our result that both phases are still present for very
large $\lambda$, where $\beta_{\mbox{\scriptsize cr}}$ becomes
negative, it is to be noted that quite some time ago a similar
phenomenon has been observed for vortex strings in the 3D O(2) spin
model \cite{ks87}. It should be further mentioned that negative
$\beta$ have also been studied in the compact U(1) lattice gauge with
complete suppression of monopoles and inclusion of matter fields, with
the observation of a first order transition at $\beta = -0.7$
\cite{hmp94,bhmp96}. Clearly this is far away from our phase
transition line which for very large $\lambda$ runs at extremely
negative $\beta$.

We have already shown in Figure 7 of Ref.~\cite{krw94} that along the
phase transition line the average plaquette energy
$E=(1/6L^4)\sum_{\mu>\nu,x} (1-\cos \Theta_{\mu\nu,x})$ increases with
$\lambda$. This increase continues for still larger $\lambda$ so that
ultimately one comes close to the upper limit of the range $0\le E \le
2$.  Thus negative plaquettes contribute significantly. In this
context it should be noted, however, that, even for $\lambda = \infty$,
for sufficiently negative $\beta$ the complete suppression of negative
plaquettes only reduces $E$ by roughly a factor 2 \cite{hmp94}.

The non-Gaussian exponent we found implies that the continuum limit of
our model is different from the usual expectations for QED and would
not be seen in perturbative approaches. Possibly monopoles survive in
the limit.  One should also be aware of the possibility of obtaining
different limits when approaching from the strong or the weak coupling
regions\cite{c81}.

As a matter of fact, we most likely have a line of critical points in
($\beta,\lambda$) space at which different continuum limits may arise
if $\nu$ is not universal. This line is to be distinguished from the
line of critical points for $\beta > \beta_{\mbox{\scriptsize cr}}$ in
case of the pure Wilson action, whose existence has been conjectured a
long time ago.  There the scale invariance of Creutz ratios \cite{b81}
hints at a diverging correlation length and thus at a line of critical
points. Similar indications come from calculations of the spectrum
\cite{bp84}.

Further work appears necessary to fully explore the critical
properties of the model and, in particular, to determine whether its
critical points can give origin to a viable theory.  A related
question is which form of the action is the most profitable to
use. The addition of a monopole term is attractive because of the close
relation of monopoles to the phase structure. We wish to note,
however, that instead of $\lambda \sum_{\rho,x} |M_{\rho,x}|$,
$\mu \sum_{\rho,x} (M_{\rho,x})^2$ or some other form might turn
out to give better results.

\section*{Acknowledgments}

\hspace{3mm}
This research was supported in part under DFG grants Ke 250/7-2 and 250/12-1
and under DOE grant DE-FG02-91ER40676.

\newpage

\renewcommand{\baselinestretch}{1.1}

\newpage

\renewcommand{\baselinestretch}{1.7}
\small\normalsize

\section*{Figure captions}

\begin{tabular}{rl}
Fig.~1. & Order parameter $P_{\mbox{\scriptsize net}}$ as function
of $\beta$ for $\lambda=0$ and $\lambda=0.9$ \\& on an $8^4$ lattice.\\

Fig.~2. & Location of the phase transition point $\beta_{\mbox{\scriptsize
cr}}$ as function of $\lambda$\\& for $8^4$ (squares) and $16^4$ (crosses)
lattices.\\

Fig.~3. & Specific heat $C$ as function of $\beta$ for $\lambda=0$,
0.9, 1.3 on an $8^4$ lattice. \\

Fig.~4. & $C_{\mbox{\scriptsize max}}$ versus $L$ for $\lambda = 0.9$
at $\beta_{\mbox{\scriptsize cr}}$.\\

Fig.~5. & Monopole density as function of $\lambda$.\\

\end{tabular}

\newpage

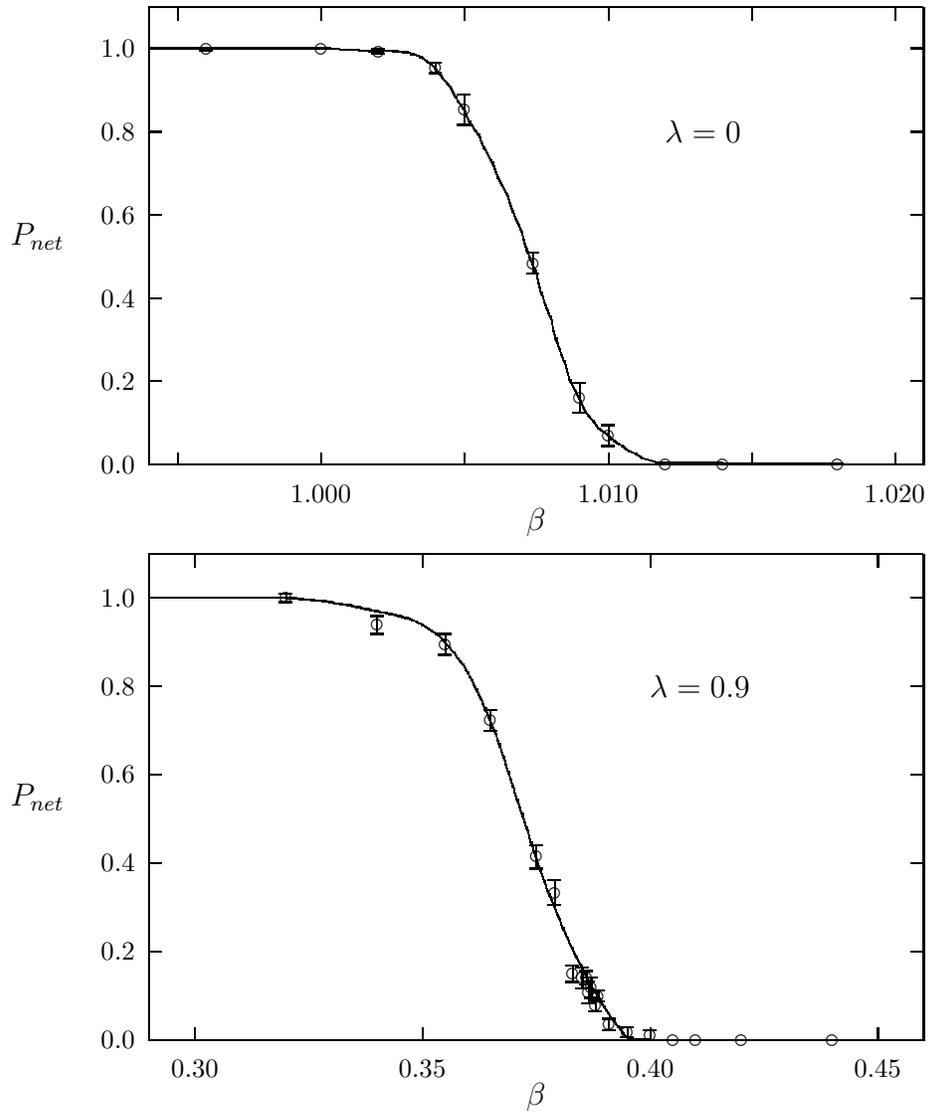
\begin{figure}[tb]
\setlength{\unitlength}{0.240900pt}
\ifx\plotpoint\undefined\newsavebox{\plotpoint}\fi
\sbox{\plotpoint}{\rule[-0.200pt]{0.400pt}{0.400pt}}%
\begin{picture}(1500,900)(0,0)
\font\gnuplot=cmr10 at 10pt
\gnuplot
\sbox{\plotpoint}{\rule[-0.200pt]{0.400pt}{0.400pt}}%
\put(220.0,113.0){\rule[-0.200pt]{4.818pt}{0.400pt}}
\put(198,113){\makebox(0,0)[r]{0.0}}
\put(1416.0,113.0){\rule[-0.200pt]{4.818pt}{0.400pt}}
\put(220.0,244.0){\rule[-0.200pt]{4.818pt}{0.400pt}}
\put(198,244){\makebox(0,0)[r]{0.2}}
\put(1416.0,244.0){\rule[-0.200pt]{4.818pt}{0.400pt}}
\put(220.0,374.0){\rule[-0.200pt]{4.818pt}{0.400pt}}
\put(198,374){\makebox(0,0)[r]{0.4}}
\put(1416.0,374.0){\rule[-0.200pt]{4.818pt}{0.400pt}}
\put(220.0,505.0){\rule[-0.200pt]{4.818pt}{0.400pt}}
\put(198,505){\makebox(0,0)[r]{0.6}}
\put(1416.0,505.0){\rule[-0.200pt]{4.818pt}{0.400pt}}
\put(220.0,636.0){\rule[-0.200pt]{4.818pt}{0.400pt}}
\put(198,636){\makebox(0,0)[r]{0.8}}
\put(1416.0,636.0){\rule[-0.200pt]{4.818pt}{0.400pt}}
\put(220.0,767.0){\rule[-0.200pt]{4.818pt}{0.400pt}}
\put(198,767){\makebox(0,0)[r]{1.0}}
\put(1416.0,767.0){\rule[-0.200pt]{4.818pt}{0.400pt}}
\put(265.0,113.0){\rule[-0.200pt]{0.400pt}{4.818pt}}
\put(265,68){\makebox(0,0){ }}
\put(265.0,812.0){\rule[-0.200pt]{0.400pt}{4.818pt}}
\put(490.0,113.0){\rule[-0.200pt]{0.400pt}{4.818pt}}
\put(490,68){\makebox(0,0){1.000}}
\put(490.0,812.0){\rule[-0.200pt]{0.400pt}{4.818pt}}
\put(715.0,113.0){\rule[-0.200pt]{0.400pt}{4.818pt}}
\put(715,68){\makebox(0,0){ }}
\put(715.0,812.0){\rule[-0.200pt]{0.400pt}{4.818pt}}
\put(941.0,113.0){\rule[-0.200pt]{0.400pt}{4.818pt}}
\put(941,68){\makebox(0,0){1.010}}
\put(941.0,812.0){\rule[-0.200pt]{0.400pt}{4.818pt}}
\put(1166.0,113.0){\rule[-0.200pt]{0.400pt}{4.818pt}}
\put(1166,68){\makebox(0,0){ }}
\put(1166.0,812.0){\rule[-0.200pt]{0.400pt}{4.818pt}}
\put(1391.0,113.0){\rule[-0.200pt]{0.400pt}{4.818pt}}
\put(1391,68){\makebox(0,0){1.020}}
\put(1391.0,812.0){\rule[-0.200pt]{0.400pt}{4.818pt}}
\put(220.0,113.0){\rule[-0.200pt]{292.934pt}{0.400pt}}
\put(1436.0,113.0){\rule[-0.200pt]{0.400pt}{173.207pt}}
\put(220.0,832.0){\rule[-0.200pt]{292.934pt}{0.400pt}}
\put(45,472){\makebox(0,0){$P_{net}$}}
\put(828,23){\makebox(0,0){$\beta$}}
\put(828,877){\makebox(0,0){ }}
\put(1031,636){\makebox(0,0)[l]{$\lambda=0$}}
\put(220.0,113.0){\rule[-0.200pt]{0.400pt}{173.207pt}}
\put(310,766){\circle{18}}
\put(490,767){\circle{18}}
\put(580,762){\circle{18}}
\put(670,736){\circle{18}}
\put(715,671){\circle{18}}
\put(823,429){\circle{18}}
\put(896,218){\circle{18}}
\put(941,159){\circle{18}}
\put(1031,113){\circle{18}}
\put(1121,114){\circle{18}}
\put(1301,113){\circle{18}}
\put(310.0,764.0){\rule[-0.200pt]{0.400pt}{0.723pt}}
\put(300.0,764.0){\rule[-0.200pt]{4.818pt}{0.400pt}}
\put(300.0,767.0){\rule[-0.200pt]{4.818pt}{0.400pt}}
\put(490.0,766.0){\rule[-0.200pt]{0.400pt}{0.482pt}}
\put(480.0,766.0){\rule[-0.200pt]{4.818pt}{0.400pt}}
\put(480.0,768.0){\rule[-0.200pt]{4.818pt}{0.400pt}}
\put(580.0,759.0){\rule[-0.200pt]{0.400pt}{1.686pt}}
\put(570.0,759.0){\rule[-0.200pt]{4.818pt}{0.400pt}}
\put(570.0,766.0){\rule[-0.200pt]{4.818pt}{0.400pt}}
\put(670.0,728.0){\rule[-0.200pt]{0.400pt}{3.854pt}}
\put(660.0,728.0){\rule[-0.200pt]{4.818pt}{0.400pt}}
\put(660.0,744.0){\rule[-0.200pt]{4.818pt}{0.400pt}}
\put(715.0,647.0){\rule[-0.200pt]{0.400pt}{11.563pt}}
\put(705.0,647.0){\rule[-0.200pt]{4.818pt}{0.400pt}}
\put(705.0,695.0){\rule[-0.200pt]{4.818pt}{0.400pt}}
\put(823.0,413.0){\rule[-0.200pt]{0.400pt}{7.950pt}}
\put(813.0,413.0){\rule[-0.200pt]{4.818pt}{0.400pt}}
\put(813.0,446.0){\rule[-0.200pt]{4.818pt}{0.400pt}}
\put(896.0,194.0){\rule[-0.200pt]{0.400pt}{11.322pt}}
\put(886.0,194.0){\rule[-0.200pt]{4.818pt}{0.400pt}}
\put(886.0,241.0){\rule[-0.200pt]{4.818pt}{0.400pt}}
\put(941.0,142.0){\rule[-0.200pt]{0.400pt}{7.950pt}}
\put(931.0,142.0){\rule[-0.200pt]{4.818pt}{0.400pt}}
\put(931.0,175.0){\rule[-0.200pt]{4.818pt}{0.400pt}}
\put(1031.0,113.0){\usebox{\plotpoint}}
\put(1021.0,113.0){\rule[-0.200pt]{4.818pt}{0.400pt}}
\put(1021.0,114.0){\rule[-0.200pt]{4.818pt}{0.400pt}}
\put(1121.0,113.0){\rule[-0.200pt]{0.400pt}{0.482pt}}
\put(1111.0,113.0){\rule[-0.200pt]{4.818pt}{0.400pt}}
\put(1111.0,115.0){\rule[-0.200pt]{4.818pt}{0.400pt}}
\put(1301.0,113.0){\usebox{\plotpoint}}
\put(1291.0,113.0){\rule[-0.200pt]{4.818pt}{0.400pt}}
\put(1291.0,114.0){\rule[-0.200pt]{4.818pt}{0.400pt}}
\multiput(490.00,765.93)(9.949,-0.477){7}{\rule{7.300pt}{0.115pt}}
\multiput(490.00,766.17)(74.848,-5.000){2}{\rule{3.650pt}{0.400pt}}
\put(580,761.67){\rule{2.168pt}{0.400pt}}
\multiput(580.00,761.17)(4.500,1.000){2}{\rule{1.084pt}{0.400pt}}
\put(220.0,767.0){\rule[-0.200pt]{65.043pt}{0.400pt}}
\put(603,761.67){\rule{2.168pt}{0.400pt}}
\multiput(603.00,762.17)(4.500,-1.000){2}{\rule{1.084pt}{0.400pt}}
\put(612,760.67){\rule{3.132pt}{0.400pt}}
\multiput(612.00,761.17)(6.500,-1.000){2}{\rule{1.566pt}{0.400pt}}
\multiput(625.00,759.95)(1.802,-0.447){3}{\rule{1.300pt}{0.108pt}}
\multiput(625.00,760.17)(6.302,-3.000){2}{\rule{0.650pt}{0.400pt}}
\multiput(634.00,756.93)(1.489,-0.477){7}{\rule{1.220pt}{0.115pt}}
\multiput(634.00,757.17)(11.468,-5.000){2}{\rule{0.610pt}{0.400pt}}
\multiput(648.00,751.93)(0.645,-0.485){11}{\rule{0.614pt}{0.117pt}}
\multiput(648.00,752.17)(7.725,-7.000){2}{\rule{0.307pt}{0.400pt}}
\multiput(657.00,744.92)(0.652,-0.491){17}{\rule{0.620pt}{0.118pt}}
\multiput(657.00,745.17)(11.713,-10.000){2}{\rule{0.310pt}{0.400pt}}
\multiput(670.59,733.19)(0.489,-0.728){15}{\rule{0.118pt}{0.678pt}}
\multiput(669.17,734.59)(9.000,-11.593){2}{\rule{0.400pt}{0.339pt}}
\multiput(679.58,720.69)(0.494,-0.570){25}{\rule{0.119pt}{0.557pt}}
\multiput(678.17,721.84)(14.000,-14.844){2}{\rule{0.400pt}{0.279pt}}
\multiput(693.59,703.45)(0.489,-0.961){15}{\rule{0.118pt}{0.856pt}}
\multiput(692.17,705.22)(9.000,-15.224){2}{\rule{0.400pt}{0.428pt}}
\multiput(702.58,687.16)(0.493,-0.734){23}{\rule{0.119pt}{0.685pt}}
\multiput(701.17,688.58)(13.000,-17.579){2}{\rule{0.400pt}{0.342pt}}
\multiput(715.59,666.90)(0.489,-1.135){15}{\rule{0.118pt}{0.989pt}}
\multiput(714.17,668.95)(9.000,-17.948){2}{\rule{0.400pt}{0.494pt}}
\multiput(724.58,648.21)(0.494,-0.717){25}{\rule{0.119pt}{0.671pt}}
\multiput(723.17,649.61)(14.000,-18.606){2}{\rule{0.400pt}{0.336pt}}
\multiput(738.59,626.71)(0.489,-1.194){15}{\rule{0.118pt}{1.033pt}}
\multiput(737.17,628.86)(9.000,-18.855){2}{\rule{0.400pt}{0.517pt}}
\multiput(747.58,606.65)(0.493,-0.893){23}{\rule{0.119pt}{0.808pt}}
\multiput(746.17,608.32)(13.000,-21.324){2}{\rule{0.400pt}{0.404pt}}
\multiput(760.59,582.16)(0.489,-1.368){15}{\rule{0.118pt}{1.167pt}}
\multiput(759.17,584.58)(9.000,-21.579){2}{\rule{0.400pt}{0.583pt}}
\multiput(769.58,559.50)(0.494,-0.938){25}{\rule{0.119pt}{0.843pt}}
\multiput(768.17,561.25)(14.000,-24.251){2}{\rule{0.400pt}{0.421pt}}
\multiput(783.59,531.60)(0.489,-1.543){15}{\rule{0.118pt}{1.300pt}}
\multiput(782.17,534.30)(9.000,-24.302){2}{\rule{0.400pt}{0.650pt}}
\multiput(792.58,505.88)(0.493,-1.131){23}{\rule{0.119pt}{0.992pt}}
\multiput(791.17,507.94)(13.000,-26.940){2}{\rule{0.400pt}{0.496pt}}
\multiput(805.59,474.68)(0.489,-1.834){15}{\rule{0.118pt}{1.522pt}}
\multiput(804.17,477.84)(9.000,-28.841){2}{\rule{0.400pt}{0.761pt}}
\multiput(814.58,444.67)(0.494,-1.195){25}{\rule{0.119pt}{1.043pt}}
\multiput(813.17,446.84)(14.000,-30.835){2}{\rule{0.400pt}{0.521pt}}
\multiput(828.59,408.94)(0.489,-2.067){15}{\rule{0.118pt}{1.700pt}}
\multiput(827.17,412.47)(9.000,-32.472){2}{\rule{0.400pt}{0.850pt}}
\multiput(837.58,375.32)(0.494,-1.305){25}{\rule{0.119pt}{1.129pt}}
\multiput(836.17,377.66)(14.000,-33.658){2}{\rule{0.400pt}{0.564pt}}
\multiput(851.59,336.94)(0.489,-2.067){15}{\rule{0.118pt}{1.700pt}}
\multiput(850.17,340.47)(9.000,-32.472){2}{\rule{0.400pt}{0.850pt}}
\multiput(860.58,303.24)(0.493,-1.329){23}{\rule{0.119pt}{1.146pt}}
\multiput(859.17,305.62)(13.000,-31.621){2}{\rule{0.400pt}{0.573pt}}
\multiput(873.59,268.05)(0.489,-1.718){15}{\rule{0.118pt}{1.433pt}}
\multiput(872.17,271.03)(9.000,-27.025){2}{\rule{0.400pt}{0.717pt}}
\multiput(882.58,240.50)(0.494,-0.938){25}{\rule{0.119pt}{0.843pt}}
\multiput(881.17,242.25)(14.000,-24.251){2}{\rule{0.400pt}{0.421pt}}
\multiput(896.59,213.71)(0.489,-1.194){15}{\rule{0.118pt}{1.033pt}}
\multiput(895.17,215.86)(9.000,-18.855){2}{\rule{0.400pt}{0.517pt}}
\multiput(905.58,194.54)(0.493,-0.616){23}{\rule{0.119pt}{0.592pt}}
\multiput(904.17,195.77)(13.000,-14.771){2}{\rule{0.400pt}{0.296pt}}
\multiput(918.59,178.37)(0.489,-0.669){15}{\rule{0.118pt}{0.633pt}}
\multiput(917.17,179.69)(9.000,-10.685){2}{\rule{0.400pt}{0.317pt}}
\multiput(927.00,167.92)(0.704,-0.491){17}{\rule{0.660pt}{0.118pt}}
\multiput(927.00,168.17)(12.630,-10.000){2}{\rule{0.330pt}{0.400pt}}
\multiput(941.00,157.93)(0.495,-0.489){15}{\rule{0.500pt}{0.118pt}}
\multiput(941.00,158.17)(7.962,-9.000){2}{\rule{0.250pt}{0.400pt}}
\multiput(950.00,148.93)(0.824,-0.488){13}{\rule{0.750pt}{0.117pt}}
\multiput(950.00,149.17)(11.443,-8.000){2}{\rule{0.375pt}{0.400pt}}
\multiput(963.00,140.93)(0.645,-0.485){11}{\rule{0.614pt}{0.117pt}}
\multiput(963.00,141.17)(7.725,-7.000){2}{\rule{0.307pt}{0.400pt}}
\multiput(972.00,133.93)(1.214,-0.482){9}{\rule{1.033pt}{0.116pt}}
\multiput(972.00,134.17)(11.855,-6.000){2}{\rule{0.517pt}{0.400pt}}
\multiput(986.00,127.93)(0.762,-0.482){9}{\rule{0.700pt}{0.116pt}}
\multiput(986.00,128.17)(7.547,-6.000){2}{\rule{0.350pt}{0.400pt}}
\multiput(995.00,121.94)(1.797,-0.468){5}{\rule{1.400pt}{0.113pt}}
\multiput(995.00,122.17)(10.094,-4.000){2}{\rule{0.700pt}{0.400pt}}
\multiput(1008.00,117.95)(1.802,-0.447){3}{\rule{1.300pt}{0.108pt}}
\multiput(1008.00,118.17)(6.302,-3.000){2}{\rule{0.650pt}{0.400pt}}
\multiput(1017.00,114.95)(83.292,-0.447){3}{\rule{49.967pt}{0.108pt}}
\multiput(1017.00,115.17)(270.292,-3.000){2}{\rule{24.983pt}{0.400pt}}
\put(589.0,763.0){\rule[-0.200pt]{3.373pt}{0.400pt}}
\end{picture}
\input fig1b.tex
\caption{}
\end{figure}
\clearpage

\begin{figure}[tb]
\input fig2.tex
\caption{}
\end{figure}
\clearpage

\begin{figure}[tb]
\setlength{\unitlength}{0.240900pt}
\ifx\plotpoint\undefined\newsavebox{\plotpoint}\fi
\begin{picture}(1500,900)(0,0)
\font\gnuplot=cmr10 at 10pt
\gnuplot
\sbox{\plotpoint}{\rule[-0.200pt]{0.400pt}{0.400pt}}%
\put(220.0,113.0){\rule[-0.200pt]{4.818pt}{0.400pt}}
\put(198,113){\makebox(0,0)[r]{0}}
\put(1416.0,113.0){\rule[-0.200pt]{4.818pt}{0.400pt}}
\put(220.0,338.0){\rule[-0.200pt]{4.818pt}{0.400pt}}
\put(198,338){\makebox(0,0)[r]{5}}
\put(1416.0,338.0){\rule[-0.200pt]{4.818pt}{0.400pt}}
\put(220.0,562.0){\rule[-0.200pt]{4.818pt}{0.400pt}}
\put(198,562){\makebox(0,0)[r]{10}}
\put(1416.0,562.0){\rule[-0.200pt]{4.818pt}{0.400pt}}
\put(220.0,787.0){\rule[-0.200pt]{4.818pt}{0.400pt}}
\put(198,787){\makebox(0,0)[r]{15}}
\put(1416.0,787.0){\rule[-0.200pt]{4.818pt}{0.400pt}}
\put(265.0,113.0){\rule[-0.200pt]{0.400pt}{4.818pt}}
\put(265,68){\makebox(0,0){ }}
\put(265.0,812.0){\rule[-0.200pt]{0.400pt}{4.818pt}}
\put(490.0,113.0){\rule[-0.200pt]{0.400pt}{4.818pt}}
\put(490,68){\makebox(0,0){1.000}}
\put(490.0,812.0){\rule[-0.200pt]{0.400pt}{4.818pt}}
\put(715.0,113.0){\rule[-0.200pt]{0.400pt}{4.818pt}}
\put(715,68){\makebox(0,0){ }}
\put(715.0,812.0){\rule[-0.200pt]{0.400pt}{4.818pt}}
\put(941.0,113.0){\rule[-0.200pt]{0.400pt}{4.818pt}}
\put(941,68){\makebox(0,0){1.010}}
\put(941.0,812.0){\rule[-0.200pt]{0.400pt}{4.818pt}}
\put(1166.0,113.0){\rule[-0.200pt]{0.400pt}{4.818pt}}
\put(1166,68){\makebox(0,0){ }}
\put(1166.0,812.0){\rule[-0.200pt]{0.400pt}{4.818pt}}
\put(1391.0,113.0){\rule[-0.200pt]{0.400pt}{4.818pt}}
\put(1391,68){\makebox(0,0){1.020}}
\put(1391.0,812.0){\rule[-0.200pt]{0.400pt}{4.818pt}}
\put(220.0,113.0){\rule[-0.200pt]{292.934pt}{0.400pt}}
\put(1436.0,113.0){\rule[-0.200pt]{0.400pt}{173.207pt}}
\put(220.0,832.0){\rule[-0.200pt]{292.934pt}{0.400pt}}
\put(45,472){\makebox(0,0){$C$}}
\put(828,23){\makebox(0,0){$\beta$}}
\put(828,877){\makebox(0,0){ }}
\put(400,652){\makebox(0,0)[l]{$\lambda=0$}}
\put(220.0,113.0){\rule[-0.200pt]{0.400pt}{173.207pt}}
\put(310,205){\makebox(0,0){$\times$}}
\put(490,218){\makebox(0,0){$\times$}}
\put(580,228){\makebox(0,0){$\times$}}
\put(1031,187){\makebox(0,0){$\times$}}
\put(1121,175){\makebox(0,0){$\times$}}
\put(670,275){\makebox(0,0){$\times$}}
\put(715,416){\makebox(0,0){$\times$}}
\put(896,481){\makebox(0,0){$\times$}}
\put(941,325){\makebox(0,0){$\times$}}
\put(1301,162){\makebox(0,0){$\times$}}
\put(823,712){\makebox(0,0){$\times$}}
\put(310.0,204.0){\rule[-0.200pt]{0.400pt}{0.482pt}}
\put(300.0,204.0){\rule[-0.200pt]{4.818pt}{0.400pt}}
\put(300.0,206.0){\rule[-0.200pt]{4.818pt}{0.400pt}}
\put(490.0,215.0){\rule[-0.200pt]{0.400pt}{1.204pt}}
\put(480.0,215.0){\rule[-0.200pt]{4.818pt}{0.400pt}}
\put(480.0,220.0){\rule[-0.200pt]{4.818pt}{0.400pt}}
\put(580.0,224.0){\rule[-0.200pt]{0.400pt}{1.927pt}}
\put(570.0,224.0){\rule[-0.200pt]{4.818pt}{0.400pt}}
\put(570.0,232.0){\rule[-0.200pt]{4.818pt}{0.400pt}}
\put(1031.0,184.0){\rule[-0.200pt]{0.400pt}{1.445pt}}
\put(1021.0,184.0){\rule[-0.200pt]{4.818pt}{0.400pt}}
\put(1021.0,190.0){\rule[-0.200pt]{4.818pt}{0.400pt}}
\put(1121.0,172.0){\rule[-0.200pt]{0.400pt}{1.445pt}}
\put(1111.0,172.0){\rule[-0.200pt]{4.818pt}{0.400pt}}
\put(1111.0,178.0){\rule[-0.200pt]{4.818pt}{0.400pt}}
\put(670.0,258.0){\rule[-0.200pt]{0.400pt}{8.191pt}}
\put(660.0,258.0){\rule[-0.200pt]{4.818pt}{0.400pt}}
\put(660.0,292.0){\rule[-0.200pt]{4.818pt}{0.400pt}}
\put(715.0,361.0){\rule[-0.200pt]{0.400pt}{26.499pt}}
\put(705.0,361.0){\rule[-0.200pt]{4.818pt}{0.400pt}}
\put(705.0,471.0){\rule[-0.200pt]{4.818pt}{0.400pt}}
\put(896.0,436.0){\rule[-0.200pt]{0.400pt}{21.922pt}}
\put(886.0,436.0){\rule[-0.200pt]{4.818pt}{0.400pt}}
\put(886.0,527.0){\rule[-0.200pt]{4.818pt}{0.400pt}}
\put(941.0,287.0){\rule[-0.200pt]{0.400pt}{18.068pt}}
\put(931.0,287.0){\rule[-0.200pt]{4.818pt}{0.400pt}}
\put(931.0,362.0){\rule[-0.200pt]{4.818pt}{0.400pt}}
\put(1301.0,161.0){\rule[-0.200pt]{0.400pt}{0.482pt}}
\put(1291.0,161.0){\rule[-0.200pt]{4.818pt}{0.400pt}}
\put(1291.0,163.0){\rule[-0.200pt]{4.818pt}{0.400pt}}
\put(823.0,696.0){\rule[-0.200pt]{0.400pt}{7.468pt}}
\put(813.0,696.0){\rule[-0.200pt]{4.818pt}{0.400pt}}
\put(813.0,727.0){\rule[-0.200pt]{4.818pt}{0.400pt}}
\end{picture}
\setlength{\unitlength}{0.240900pt}
\ifx\plotpoint\undefined\newsavebox{\plotpoint}\fi
\begin{picture}(1500,900)(0,0)
\font\gnuplot=cmr10 at 10pt
\gnuplot
\sbox{\plotpoint}{\rule[-0.200pt]{0.400pt}{0.400pt}}%
\put(220.0,113.0){\rule[-0.200pt]{4.818pt}{0.400pt}}
\put(198,113){\makebox(0,0)[r]{0.5}}
\put(1416.0,113.0){\rule[-0.200pt]{4.818pt}{0.400pt}}
\put(220.0,473.0){\rule[-0.200pt]{4.818pt}{0.400pt}}
\put(198,473){\makebox(0,0)[r]{1.0}}
\put(1416.0,473.0){\rule[-0.200pt]{4.818pt}{0.400pt}}
\put(220.0,832.0){\rule[-0.200pt]{4.818pt}{0.400pt}}
\put(198,832){\makebox(0,0)[r]{1.5}}
\put(1416.0,832.0){\rule[-0.200pt]{4.818pt}{0.400pt}}
\put(314.0,113.0){\rule[-0.200pt]{0.400pt}{4.818pt}}
\put(314,68){\makebox(0,0){ }}
\put(314.0,812.0){\rule[-0.200pt]{0.400pt}{4.818pt}}
\put(547.0,113.0){\rule[-0.200pt]{0.400pt}{4.818pt}}
\put(547,68){\makebox(0,0){0.35}}
\put(547.0,812.0){\rule[-0.200pt]{0.400pt}{4.818pt}}
\put(781.0,113.0){\rule[-0.200pt]{0.400pt}{4.818pt}}
\put(781,68){\makebox(0,0){ }}
\put(781.0,812.0){\rule[-0.200pt]{0.400pt}{4.818pt}}
\put(1015.0,113.0){\rule[-0.200pt]{0.400pt}{4.818pt}}
\put(1015,68){\makebox(0,0){0.40}}
\put(1015.0,812.0){\rule[-0.200pt]{0.400pt}{4.818pt}}
\put(1249.0,113.0){\rule[-0.200pt]{0.400pt}{4.818pt}}
\put(1249,68){\makebox(0,0){ }}
\put(1249.0,812.0){\rule[-0.200pt]{0.400pt}{4.818pt}}
\put(220.0,113.0){\rule[-0.200pt]{292.934pt}{0.400pt}}
\put(1436.0,113.0){\rule[-0.200pt]{0.400pt}{173.207pt}}
\put(220.0,832.0){\rule[-0.200pt]{292.934pt}{0.400pt}}
\put(45,472){\makebox(0,0){$C$}}
\put(828,23){\makebox(0,0){$\beta$}}
\put(828,877){\makebox(0,0){ }}
\put(360,652){\makebox(0,0)[l]{$\lambda=0.9$}}
\put(220.0,113.0){\rule[-0.200pt]{0.400pt}{173.207pt}}
\put(875,664){\makebox(0,0){$\times$}}
\put(884,714){\makebox(0,0){$\times$}}
\put(889,710){\makebox(0,0){$\times$}}
\put(893,733){\makebox(0,0){$\times$}}
\put(903,748){\makebox(0,0){$\times$}}
\put(856,577){\makebox(0,0){$\times$}}
\put(968,598){\makebox(0,0){$\times$}}
\put(594,382){\makebox(0,0){$\times$}}
\put(688,421){\makebox(0,0){$\times$}}
\put(908,714){\makebox(0,0){$\times$}}
\put(1015,394){\makebox(0,0){$\times$}}
\put(267,302){\makebox(0,0){$\times$}}
\put(454,357){\makebox(0,0){$\times$}}
\put(1062,319){\makebox(0,0){$\times$}}
\put(1202,252){\makebox(0,0){$\times$}}
\put(1389,216){\makebox(0,0){$\times$}}
\put(875.0,643.0){\rule[-0.200pt]{0.400pt}{10.359pt}}
\put(865.0,643.0){\rule[-0.200pt]{4.818pt}{0.400pt}}
\put(865.0,686.0){\rule[-0.200pt]{4.818pt}{0.400pt}}
\put(884.0,671.0){\rule[-0.200pt]{0.400pt}{20.717pt}}
\put(874.0,671.0){\rule[-0.200pt]{4.818pt}{0.400pt}}
\put(874.0,757.0){\rule[-0.200pt]{4.818pt}{0.400pt}}
\put(889.0,681.0){\rule[-0.200pt]{0.400pt}{13.972pt}}
\put(879.0,681.0){\rule[-0.200pt]{4.818pt}{0.400pt}}
\put(879.0,739.0){\rule[-0.200pt]{4.818pt}{0.400pt}}
\put(893.0,711.0){\rule[-0.200pt]{0.400pt}{10.359pt}}
\put(883.0,711.0){\rule[-0.200pt]{4.818pt}{0.400pt}}
\put(883.0,754.0){\rule[-0.200pt]{4.818pt}{0.400pt}}
\put(903.0,726.0){\rule[-0.200pt]{0.400pt}{10.359pt}}
\put(893.0,726.0){\rule[-0.200pt]{4.818pt}{0.400pt}}
\put(893.0,769.0){\rule[-0.200pt]{4.818pt}{0.400pt}}
\put(856.0,541.0){\rule[-0.200pt]{0.400pt}{17.345pt}}
\put(846.0,541.0){\rule[-0.200pt]{4.818pt}{0.400pt}}
\put(846.0,613.0){\rule[-0.200pt]{4.818pt}{0.400pt}}
\put(968.0,547.0){\rule[-0.200pt]{0.400pt}{24.331pt}}
\put(958.0,547.0){\rule[-0.200pt]{4.818pt}{0.400pt}}
\put(958.0,648.0){\rule[-0.200pt]{4.818pt}{0.400pt}}
\put(594.0,368.0){\rule[-0.200pt]{0.400pt}{6.745pt}}
\put(584.0,368.0){\rule[-0.200pt]{4.818pt}{0.400pt}}
\put(584.0,396.0){\rule[-0.200pt]{4.818pt}{0.400pt}}
\put(688.0,404.0){\rule[-0.200pt]{0.400pt}{8.432pt}}
\put(678.0,404.0){\rule[-0.200pt]{4.818pt}{0.400pt}}
\put(678.0,439.0){\rule[-0.200pt]{4.818pt}{0.400pt}}
\put(908.0,694.0){\rule[-0.200pt]{0.400pt}{9.636pt}}
\put(898.0,694.0){\rule[-0.200pt]{4.818pt}{0.400pt}}
\put(898.0,734.0){\rule[-0.200pt]{4.818pt}{0.400pt}}
\put(1015.0,358.0){\rule[-0.200pt]{0.400pt}{17.345pt}}
\put(1005.0,358.0){\rule[-0.200pt]{4.818pt}{0.400pt}}
\put(1005.0,430.0){\rule[-0.200pt]{4.818pt}{0.400pt}}
\put(267.0,288.0){\rule[-0.200pt]{0.400pt}{6.745pt}}
\put(257.0,288.0){\rule[-0.200pt]{4.818pt}{0.400pt}}
\put(257.0,316.0){\rule[-0.200pt]{4.818pt}{0.400pt}}
\put(454.0,342.0){\rule[-0.200pt]{0.400pt}{6.986pt}}
\put(444.0,342.0){\rule[-0.200pt]{4.818pt}{0.400pt}}
\put(444.0,371.0){\rule[-0.200pt]{4.818pt}{0.400pt}}
\put(1062.0,306.0){\rule[-0.200pt]{0.400pt}{6.504pt}}
\put(1052.0,306.0){\rule[-0.200pt]{4.818pt}{0.400pt}}
\put(1052.0,333.0){\rule[-0.200pt]{4.818pt}{0.400pt}}
\put(1202.0,241.0){\rule[-0.200pt]{0.400pt}{5.300pt}}
\put(1192.0,241.0){\rule[-0.200pt]{4.818pt}{0.400pt}}
\put(1192.0,263.0){\rule[-0.200pt]{4.818pt}{0.400pt}}
\put(1389.0,209.0){\rule[-0.200pt]{0.400pt}{3.373pt}}
\put(1379.0,209.0){\rule[-0.200pt]{4.818pt}{0.400pt}}
\put(1379.0,223.0){\rule[-0.200pt]{4.818pt}{0.400pt}}
\end{picture}
\setlength{\unitlength}{0.240900pt}
\ifx\plotpoint\undefined\newsavebox{\plotpoint}\fi
\begin{picture}(1500,900)(0,0)
\font\gnuplot=cmr10 at 10pt
\gnuplot
\sbox{\plotpoint}{\rule[-0.200pt]{0.400pt}{0.400pt}}%
\put(220.0,293.0){\rule[-0.200pt]{4.818pt}{0.400pt}}
\put(198,293){\makebox(0,0)[r]{0.65}}
\put(1416.0,293.0){\rule[-0.200pt]{4.818pt}{0.400pt}}
\put(220.0,517.0){\rule[-0.200pt]{4.818pt}{0.400pt}}
\put(198,517){\makebox(0,0)[r]{0.675}}
\put(1416.0,517.0){\rule[-0.200pt]{4.818pt}{0.400pt}}
\put(220.0,742.0){\rule[-0.200pt]{4.818pt}{0.400pt}}
\put(198,742){\makebox(0,0)[r]{0.70}}
\put(1416.0,742.0){\rule[-0.200pt]{4.818pt}{0.400pt}}
\put(372.0,113.0){\rule[-0.200pt]{0.400pt}{4.818pt}}
\put(372,68){\makebox(0,0){-0.4}}
\put(372.0,812.0){\rule[-0.200pt]{0.400pt}{4.818pt}}
\put(676.0,113.0){\rule[-0.200pt]{0.400pt}{4.818pt}}
\put(676,68){\makebox(0,0){-0.3}}
\put(676.0,812.0){\rule[-0.200pt]{0.400pt}{4.818pt}}
\put(980.0,113.0){\rule[-0.200pt]{0.400pt}{4.818pt}}
\put(980,68){\makebox(0,0){-0.2}}
\put(980.0,812.0){\rule[-0.200pt]{0.400pt}{4.818pt}}
\put(1284.0,113.0){\rule[-0.200pt]{0.400pt}{4.818pt}}
\put(1284,68){\makebox(0,0){-0.1}}
\put(1284.0,812.0){\rule[-0.200pt]{0.400pt}{4.818pt}}
\put(220.0,113.0){\rule[-0.200pt]{292.934pt}{0.400pt}}
\put(1436.0,113.0){\rule[-0.200pt]{0.400pt}{173.207pt}}
\put(220.0,832.0){\rule[-0.200pt]{292.934pt}{0.400pt}}
\put(45,472){\makebox(0,0){$C$}}
\put(828,23){\makebox(0,0){$\beta$}}
\put(828,877){\makebox(0,0){ }}
\put(433,607){\makebox(0,0)[l]{$\lambda=1.3$}}
\put(220.0,113.0){\rule[-0.200pt]{0.400pt}{173.207pt}}
\put(372,293){\makebox(0,0){$\times$}}
\put(676,293){\makebox(0,0){$\times$}}
\put(828,293){\makebox(0,0){$\times$}}
\put(950,473){\makebox(0,0){$\times$}}
\put(980,580){\makebox(0,0){$\times$}}
\put(995,697){\makebox(0,0){$\times$}}
\put(1010,589){\makebox(0,0){$\times$}}
\put(1132,517){\makebox(0,0){$\times$}}
\put(1284,239){\makebox(0,0){$\times$}}
\put(372.0,248.0){\rule[-0.200pt]{0.400pt}{21.681pt}}
\put(362.0,248.0){\rule[-0.200pt]{4.818pt}{0.400pt}}
\put(362.0,338.0){\rule[-0.200pt]{4.818pt}{0.400pt}}
\put(676.0,248.0){\rule[-0.200pt]{0.400pt}{21.681pt}}
\put(666.0,248.0){\rule[-0.200pt]{4.818pt}{0.400pt}}
\put(666.0,338.0){\rule[-0.200pt]{4.818pt}{0.400pt}}
\put(828.0,248.0){\rule[-0.200pt]{0.400pt}{21.681pt}}
\put(818.0,248.0){\rule[-0.200pt]{4.818pt}{0.400pt}}
\put(818.0,338.0){\rule[-0.200pt]{4.818pt}{0.400pt}}
\put(950.0,428.0){\rule[-0.200pt]{0.400pt}{21.440pt}}
\put(940.0,428.0){\rule[-0.200pt]{4.818pt}{0.400pt}}
\put(940.0,517.0){\rule[-0.200pt]{4.818pt}{0.400pt}}
\put(980.0,508.0){\rule[-0.200pt]{0.400pt}{34.690pt}}
\put(970.0,508.0){\rule[-0.200pt]{4.818pt}{0.400pt}}
\put(970.0,652.0){\rule[-0.200pt]{4.818pt}{0.400pt}}
\put(995.0,625.0){\rule[-0.200pt]{0.400pt}{34.690pt}}
\put(985.0,625.0){\rule[-0.200pt]{4.818pt}{0.400pt}}
\put(985.0,769.0){\rule[-0.200pt]{4.818pt}{0.400pt}}
\put(1010.0,517.0){\rule[-0.200pt]{0.400pt}{34.690pt}}
\put(1000.0,517.0){\rule[-0.200pt]{4.818pt}{0.400pt}}
\put(1000.0,661.0){\rule[-0.200pt]{4.818pt}{0.400pt}}
\put(1132.0,473.0){\rule[-0.200pt]{0.400pt}{21.440pt}}
\put(1122.0,473.0){\rule[-0.200pt]{4.818pt}{0.400pt}}
\put(1122.0,562.0){\rule[-0.200pt]{4.818pt}{0.400pt}}
\put(1284.0,194.0){\rule[-0.200pt]{0.400pt}{21.681pt}}
\put(1274.0,194.0){\rule[-0.200pt]{4.818pt}{0.400pt}}
\put(1274.0,284.0){\rule[-0.200pt]{4.818pt}{0.400pt}}
\end{picture}
\caption{}
\end{figure}
\clearpage

\begin{figure}[tb]
\setlength{\unitlength}{0.240900pt}
\ifx\plotpoint\undefined\newsavebox{\plotpoint}\fi
\begin{picture}(1500,900)(0,0)
\font\gnuplot=cmr10 at 10pt
\gnuplot
\sbox{\plotpoint}{\rule[-0.200pt]{0.400pt}{0.400pt}}%
\put(220.0,231.0){\rule[-0.200pt]{4.818pt}{0.400pt}}
\put(198,231){\makebox(0,0)[r]{1.2}}
\put(1416.0,231.0){\rule[-0.200pt]{4.818pt}{0.400pt}}
\put(220.0,453.0){\rule[-0.200pt]{4.818pt}{0.400pt}}
\put(198,453){\makebox(0,0)[r]{1.4}}
\put(1416.0,453.0){\rule[-0.200pt]{4.818pt}{0.400pt}}
\put(220.0,645.0){\rule[-0.200pt]{4.818pt}{0.400pt}}
\put(198,645){\makebox(0,0)[r]{1.6}}
\put(1416.0,645.0){\rule[-0.200pt]{4.818pt}{0.400pt}}
\put(220.0,814.0){\rule[-0.200pt]{4.818pt}{0.400pt}}
\put(198,814){\makebox(0,0)[r]{1.8}}
\put(1416.0,814.0){\rule[-0.200pt]{4.818pt}{0.400pt}}
\put(384.0,113.0){\rule[-0.200pt]{0.400pt}{4.818pt}}
\put(384,68){\makebox(0,0){6}}
\put(384.0,812.0){\rule[-0.200pt]{0.400pt}{4.818pt}}
\put(717.0,113.0){\rule[-0.200pt]{0.400pt}{4.818pt}}
\put(717,68){\makebox(0,0){8}}
\put(717.0,812.0){\rule[-0.200pt]{0.400pt}{4.818pt}}
\put(976.0,113.0){\rule[-0.200pt]{0.400pt}{4.818pt}}
\put(976,68){\makebox(0,0){10}}
\put(976.0,812.0){\rule[-0.200pt]{0.400pt}{4.818pt}}
\put(1187.0,113.0){\rule[-0.200pt]{0.400pt}{4.818pt}}
\put(1187,68){\makebox(0,0){12}}
\put(1187.0,812.0){\rule[-0.200pt]{0.400pt}{4.818pt}}
\put(220.0,113.0){\rule[-0.200pt]{292.934pt}{0.400pt}}
\put(1436.0,113.0){\rule[-0.200pt]{0.400pt}{173.207pt}}
\put(220.0,832.0){\rule[-0.200pt]{292.934pt}{0.400pt}}
\put(45,472){\makebox(0,0){$C_{max}$}}
\put(828,23){\makebox(0,0){$L$}}
\put(828,877){\makebox(0,0){ }}
\put(220.0,113.0){\rule[-0.200pt]{0.400pt}{173.207pt}}
\put(384,219){\makebox(0,0){$\times$}}
\put(717,412){\makebox(0,0){$\times$}}
\put(976,552){\makebox(0,0){$\times$}}
\put(1187,703){\makebox(0,0){$\times$}}
\put(384.0,195.0){\rule[-0.200pt]{0.400pt}{11.804pt}}
\put(374.0,244.0){\rule[-0.200pt]{4.818pt}{0.400pt}}
\put(374.0,195.0){\rule[-0.200pt]{4.818pt}{0.400pt}}
\put(717.0,379.0){\rule[-0.200pt]{0.400pt}{15.177pt}}
\put(707.0,442.0){\rule[-0.200pt]{4.818pt}{0.400pt}}
\put(707.0,379.0){\rule[-0.200pt]{4.818pt}{0.400pt}}
\put(976.0,523.0){\rule[-0.200pt]{0.400pt}{13.731pt}}
\put(966.0,580.0){\rule[-0.200pt]{4.818pt}{0.400pt}}
\put(966.0,523.0){\rule[-0.200pt]{4.818pt}{0.400pt}}
\put(1187.0,677.0){\rule[-0.200pt]{0.400pt}{13.009pt}}
\put(1177.0,731.0){\rule[-0.200pt]{4.818pt}{0.400pt}}
\put(1177.0,677.0){\rule[-0.200pt]{4.818pt}{0.400pt}}
\sbox{\plotpoint}{\rule[-0.400pt]{0.800pt}{0.800pt}}%
\put(278,149){\usebox{\plotpoint}}
\multiput(278.00,150.41)(0.842,0.500){1231}{\rule{1.547pt}{0.120pt}}
\multiput(278.00,147.34)(1038.790,619.000){2}{\rule{0.773pt}{0.800pt}}
\end{picture}
\caption{}
\end{figure}

\begin{figure}[tb]
\setlength{\unitlength}{0.240900pt}
\ifx\plotpoint\undefined\newsavebox{\plotpoint}\fi
\begin{picture}(1500,900)(0,0)
\font\gnuplot=cmr10 at 10pt
\gnuplot
\sbox{\plotpoint}{\rule[-0.200pt]{0.400pt}{0.400pt}}%
\put(220.0,712.0){\rule[-0.200pt]{4.818pt}{0.400pt}}
\put(198,712){\makebox(0,0)[r]{0.08}}
\put(1416.0,712.0){\rule[-0.200pt]{4.818pt}{0.400pt}}
\put(220.0,473.0){\rule[-0.200pt]{4.818pt}{0.400pt}}
\put(198,473){\makebox(0,0)[r]{0.06}}
\put(1416.0,473.0){\rule[-0.200pt]{4.818pt}{0.400pt}}
\put(220.0,233.0){\rule[-0.200pt]{4.818pt}{0.400pt}}
\put(198,233){\makebox(0,0)[r]{0.04}}
\put(1416.0,233.0){\rule[-0.200pt]{4.818pt}{0.400pt}}
\put(220.0,113.0){\rule[-0.200pt]{0.400pt}{4.818pt}}
\put(220,68){\makebox(0,0){-0.5}}
\put(220.0,812.0){\rule[-0.200pt]{0.400pt}{4.818pt}}
\put(524.0,113.0){\rule[-0.200pt]{0.400pt}{4.818pt}}
\put(524,68){\makebox(0,0){0.0}}
\put(524.0,812.0){\rule[-0.200pt]{0.400pt}{4.818pt}}
\put(828.0,113.0){\rule[-0.200pt]{0.400pt}{4.818pt}}
\put(828,68){\makebox(0,0){0.5}}
\put(828.0,812.0){\rule[-0.200pt]{0.400pt}{4.818pt}}
\put(1132.0,113.0){\rule[-0.200pt]{0.400pt}{4.818pt}}
\put(1132,68){\makebox(0,0){1.0}}
\put(1132.0,812.0){\rule[-0.200pt]{0.400pt}{4.818pt}}
\put(1436.0,113.0){\rule[-0.200pt]{0.400pt}{4.818pt}}
\put(1436,68){\makebox(0,0){1.5}}
\put(1436.0,812.0){\rule[-0.200pt]{0.400pt}{4.818pt}}
\put(220.0,113.0){\rule[-0.200pt]{292.934pt}{0.400pt}}
\put(1436.0,113.0){\rule[-0.200pt]{0.400pt}{173.207pt}}
\put(220.0,832.0){\rule[-0.200pt]{292.934pt}{0.400pt}}
\put(45,472){\makebox(0,0){$\rho$}}
\put(828,23){\makebox(0,0){$\lambda$}}
\put(828,877){\makebox(0,0){ }}
\put(950,12935){\makebox(0,0)[l]{ }}
\put(220.0,113.0){\rule[-0.200pt]{0.400pt}{173.207pt}}
\put(342,760){\makebox(0,0){$\times$}}
\put(402,692){\makebox(0,0){$\times$}}
\put(463,632){\makebox(0,0){$\times$}}
\put(524,580){\makebox(0,0){$\times$}}
\put(585,526){\makebox(0,0){$\times$}}
\put(646,481){\makebox(0,0){$\times$}}
\put(706,437){\makebox(0,0){$\times$}}
\put(767,390){\makebox(0,0){$\times$}}
\put(828,347){\makebox(0,0){$\times$}}
\put(889,314){\makebox(0,0){$\times$}}
\put(1010,269){\makebox(0,0){$\times$}}
\put(1071,245){\makebox(0,0){$\times$}}
\put(1132,245){\makebox(0,0){$\times$}}
\put(1193,239){\makebox(0,0){$\times$}}
\put(1254,245){\makebox(0,0){$\times$}}
\put(1314,245){\makebox(0,0){$\times$}}
\put(342.0,748.0){\rule[-0.200pt]{0.400pt}{5.782pt}}
\put(332.0,748.0){\rule[-0.200pt]{4.818pt}{0.400pt}}
\put(332.0,772.0){\rule[-0.200pt]{4.818pt}{0.400pt}}
\put(402.0,680.0){\rule[-0.200pt]{0.400pt}{5.782pt}}
\put(392.0,680.0){\rule[-0.200pt]{4.818pt}{0.400pt}}
\put(392.0,704.0){\rule[-0.200pt]{4.818pt}{0.400pt}}
\put(463.0,620.0){\rule[-0.200pt]{0.400pt}{5.782pt}}
\put(453.0,620.0){\rule[-0.200pt]{4.818pt}{0.400pt}}
\put(453.0,644.0){\rule[-0.200pt]{4.818pt}{0.400pt}}
\put(524.0,568.0){\rule[-0.200pt]{0.400pt}{5.782pt}}
\put(514.0,568.0){\rule[-0.200pt]{4.818pt}{0.400pt}}
\put(514.0,592.0){\rule[-0.200pt]{4.818pt}{0.400pt}}
\put(585.0,514.0){\rule[-0.200pt]{0.400pt}{5.782pt}}
\put(575.0,514.0){\rule[-0.200pt]{4.818pt}{0.400pt}}
\put(575.0,538.0){\rule[-0.200pt]{4.818pt}{0.400pt}}
\put(646.0,469.0){\rule[-0.200pt]{0.400pt}{5.782pt}}
\put(636.0,469.0){\rule[-0.200pt]{4.818pt}{0.400pt}}
\put(636.0,493.0){\rule[-0.200pt]{4.818pt}{0.400pt}}
\put(706.0,425.0){\rule[-0.200pt]{0.400pt}{5.782pt}}
\put(696.0,425.0){\rule[-0.200pt]{4.818pt}{0.400pt}}
\put(696.0,449.0){\rule[-0.200pt]{4.818pt}{0.400pt}}
\put(767.0,378.0){\rule[-0.200pt]{0.400pt}{5.782pt}}
\put(757.0,378.0){\rule[-0.200pt]{4.818pt}{0.400pt}}
\put(757.0,402.0){\rule[-0.200pt]{4.818pt}{0.400pt}}
\put(828.0,335.0){\rule[-0.200pt]{0.400pt}{5.782pt}}
\put(818.0,335.0){\rule[-0.200pt]{4.818pt}{0.400pt}}
\put(818.0,359.0){\rule[-0.200pt]{4.818pt}{0.400pt}}
\put(889.0,302.0){\rule[-0.200pt]{0.400pt}{5.782pt}}
\put(879.0,302.0){\rule[-0.200pt]{4.818pt}{0.400pt}}
\put(879.0,326.0){\rule[-0.200pt]{4.818pt}{0.400pt}}
\put(1010.0,254.0){\rule[-0.200pt]{0.400pt}{6.986pt}}
\put(1000.0,254.0){\rule[-0.200pt]{4.818pt}{0.400pt}}
\put(1000.0,283.0){\rule[-0.200pt]{4.818pt}{0.400pt}}
\put(1071.0,230.0){\rule[-0.200pt]{0.400pt}{6.986pt}}
\put(1061.0,230.0){\rule[-0.200pt]{4.818pt}{0.400pt}}
\put(1061.0,259.0){\rule[-0.200pt]{4.818pt}{0.400pt}}
\put(1132.0,230.0){\rule[-0.200pt]{0.400pt}{6.986pt}}
\put(1122.0,230.0){\rule[-0.200pt]{4.818pt}{0.400pt}}
\put(1122.0,259.0){\rule[-0.200pt]{4.818pt}{0.400pt}}
\put(1193.0,222.0){\rule[-0.200pt]{0.400pt}{8.191pt}}
\put(1183.0,222.0){\rule[-0.200pt]{4.818pt}{0.400pt}}
\put(1183.0,256.0){\rule[-0.200pt]{4.818pt}{0.400pt}}
\put(1254.0,228.0){\rule[-0.200pt]{0.400pt}{8.191pt}}
\put(1244.0,228.0){\rule[-0.200pt]{4.818pt}{0.400pt}}
\put(1244.0,262.0){\rule[-0.200pt]{4.818pt}{0.400pt}}
\put(1314.0,228.0){\rule[-0.200pt]{0.400pt}{8.191pt}}
\put(1304.0,228.0){\rule[-0.200pt]{4.818pt}{0.400pt}}
\put(1304.0,262.0){\rule[-0.200pt]{4.818pt}{0.400pt}}
\end{picture}
\caption{}
\end{figure}

\end{document}